\documentclass[12pt,preprint]{aastex}

\begin{document}
\title{Radio Frequency Spectra of 388 Bright 74 MHz Sources}
\author{J. F. Helmboldt\altaffilmark{1}}
\author{N. E. Kassim\altaffilmark{1}}
\author{A. S. Cohen\altaffilmark{1}}
\author{W. M. Lane\altaffilmark{1}}
\author{T. J. Lazio\altaffilmark{1}}
\email{joe.helmboldt@nrl.navy.mil}

\altaffiltext{1}{Naval Research Laboratory, Code 7213, 4555 Overlook Avenue SW, Washington, DC 20375-5351}


\begin{abstract}
As a service to the community, we have compiled radio frequency spectra from the literature for all  sources within the VLA Low Frequency Sky Survey (VLSS) that are brighter than 15 Jy at 74 MHz.  Over 160 references were used to maximize the amount of spectral data used in the compilation of the spectra, while also taking care to determine the corrections needed to put the flux densities from all reference on the same absolute flux density scale.  With the new VLSS data, we are able to vastly improve upon previous efforts to compile spectra of bright radio sources to frequencies below 100 MHz because (1) the VLSS flux densities are more reliable than those from some previous low frequency surveys and (2) the VLSS covers a much larger area of the sky ($\delta >-30^{\circ}$) than many other low frequency surveys (e.g., the 8C survey).  In this paper, we discuss how the spectra were constructed and how parameters quantifying the shapes of the spectra were derived.  Both the spectra and the shape parameters are made available here to assist in the calibration of observations made with current and future low frequency radio facilities.

\end{abstract}

\keywords{catalogs --- surveys --- radio continuum: general}

\section{Introduction}
With the implementation of the 74 MHz system for the Very Large Array \citep[VLA; ][]{kas07} and the development of the Long Wavelength Array \citep[LWA; ][]{kas06} and the Low Frequency Array \citep[LOFAR; ][]{kas04}, a new wavelength regime is becoming available for extragalactic astronomy.  The spatial resolution of the 74 MHz VLA system \citep[$\sim$25 arcsec, and as good as $\sim$10 arcsec with the Pie Town link; ][]{laz06} is vastly superior to previous low frequency instruments (e.g. the Clark Lake Radio Observatory TPT array had a resolution of 74 arcmin at 5 MHz); the LWA and LOFAR will have even better resolution (approaching 1 arcsec at 74 MHz).  This will allow for spatially resolved studies of relatively nearby galaxies and large surveys of extragalactic sources with relatively small angular sizes at low frequencies that were not possible before.\par
An important first step into this new area of extragalactic astronomy is the VLA Low-frequency Sky Survey \citep[VLSS; ][]{coh07}. The VLSS is a nearly completed survey of the northern sky (declinations$>-30^{\circ}$) with the VLA at 74 MHz with 80 arcsec resolution and an rms noise $\sim$100 mJy.  From the VLSS observations, an image database and a catalog of $\sim$68,000 sources have been constructed and made publicly available\footnote{http://lwa.nrl.navy.mil/VLSS}.  For the vast majority of sources within the VLSS catalog, the VLSS 74 MHz flux density represents the best and/or only low-frequency measurement available.  This, along with an extensive literature search of the latest available flux density measurements at other frequencies, has allowed us to extend reliably the spectra of most major radio sources to below 100 MHz for the first time.  These data will provide a better understanding of the nature of extragalactic radio sources.  The data will also useful for interpolating the flux densities of bright radio sources at low and intermediate frequencies for use as calibrators for the next generation of low frequency radio telescopes now being developed.  In this paper, we discuss how the spectra were obtained and by what means we have chosen to characterize their shapes.

\section{The Spectra}
\subsection{Compilation of Spectral Data}
In order to produce a sample of low frequency ($<$100 MHz) sources which are likely to have been previously observed and detected at other frequencies, we have limited ourselves to the brightest VLSS sources.  To ensure that the 74 MHz VLSS flux densities of these sources are accurate, we have chosen a peak brightness limit of 15 Jy beam$^{-1}$ at 74 MHz.  For about 95\% of the sources in the VLSS catalog, this limit is at least 100 times the local rms of the VLSS images and is $\sim$100 times larger than the estimated magnitude of any systematic bias(es) \citep[e.g., clean bias; see ][]{con98} in the measured flux densities \citep{coh07}.\par
We have also further refined the source list within the VLSS catalog to account for sources with relatively large angular sizes and significant amounts of structure.  Since the VLSS catalog was constructed by fitting Gaussian components to significant ($>5\sigma$) peaks in the images \citep[see ][]{coh07}, it is possible that such a source was broken up and its components were cataloged separately.  To compensate for this, we have searched the catalog for groups of Gaussian components that overlap on the sky.  For this purpose, the outer boundary of each component was defined to be an ellipse with the same position and position angle as the Gaussian fit and which extends 3$\sigma$ from the center (i.e., with major and minor axes equal to $3a/\sqrt{8\mbox{ln}2}$ and $3b/\sqrt{8\mbox{ln}2}$ where $a$ and $b$ and the full widths at half maximum of the Gaussian fit along the major and minor axes, respectively).  Components that had outer boundaries that were defined in this manner which intersected were considered to be overlapping and components of the same source.  For each of these groups, the position and peak brightness of the group was taken to be the position and peak brightness of the component with the largest peak brightness.  The components of these groups were not considered as separate sources for the work presented here.  The largest of these multi-component source have 4$-$7 components and angular sizes of $\sim$1 arcminute.  On average, the unaltered VLSS catalog contains about 2.2 sources per square degree, making it unlikely that a significant number of these multi-component sources are caused by chance projections of physically unrelated sources.\par
There are 388 VLSS sources with peak 74 MHz brightnesses $>$15 Jy beam$^{-1}$, 25 of which have multiple components according to the above definition.  To begin our construction of the radio frequency spectra of these 388 sources, we have chosen to search within five major radio surveys to obtain flux density measurements for these sources at roughly 300, 1400, and 5000 MHz.  By using these well known radio surveys which have been used extensively in the literature and which have proven to provide reliable flux densities, we will ensue that each spectrum has a set of "anchor points".  These anchor points will help constrain the shape of each spectrum for intermediate to higher frequencies, even when more data from the literature are added at other frequencies (see below) which may be less reliable.  To cover the entire range in declination spanned by the VLSS, we have searched both the Westerbork Northern Sky Survey \citep[WENSS; 327 MHz; ][]{ren97}, which covers the northern sky above a declination of 30$^{\circ}$ and the Texas Survey of Radio Sources \citep[TXS; 365 MHz; ][]{dou96}, which spans a declination range between $-35^{\circ}$ and $71.5^{\circ}$.  Similarly, in order to obtain 5000 MHz flux densities, we have searched both the Parkes-MIT-NRAO Sky Surveys \citep[PMN; 4850 MHz; ][]{wri94,wri96,gri95} and the GB6 Catalog of Radio Sources \citep[GB6; 4850 MHz; ][]{gre96}, which cover the declination ranges of $-87.5^{\circ}<\delta<10^{\circ}$ and $0^{\circ}<\delta<75^{\circ}$, respectively.  For the 1400 MHz flux densities, we have chosen the NRAO VLA Sky Survey \citep[NVSSS; ][]{con98} which covers all declinations above $-40^{\circ}$.  We have searched all five catalogs for matches to VLSS sources with a search radius of 40 arcsec (i.e., one half the FWHM of the VLSS beam).  For each source with multiple components, the search was performed separately for each component and the flux densities from any matches were added together.  For any source with matches to sources from both the WENSS and TXS catalogs, the flux density of the WENSS source was used.  For any source with matches to sources from both the PMN and GB6 catalogs, an average of the flux densities from the two matches was used.  For sources with declinations $>75^{\circ}$ that are beyond the areas covered by the PMN and GB6 catalogs, other references from the literature were used for the 5000 MHz flux density (see below).  The basic properties of these five catalogs are summarized in Table \ref{mreftab}.  The flux densities from all of these surveys except TXS are on the absolute flux density scale of \citet{baa77}, as are the VLSS flux densities.  To put the TXS flux densities onto the \citet{baa77} scale, we multiplied them by a factor of 1.041 \citep{dou96}. \par
In order to obtain additional spectral data for each source, we have also performed a more thorough search of the literature using the Astrophysical CATalogs support System (CATS) database\footnote{http://cats.sao.ru} \citep{ver97}.  The CATS database provides a searchable compilation of radio flux densities for sources taken from over 350 different references.  From this list of references, we have excluded those that are compilations of flux densities from the literature and contain no new data as well as those that used very long baseline interferometry (VLBI), as such observations may miss a significant amount of the total flux density of a source.  For each reference with at least one match to a VLSS source, we determined the multiplicative factor needed to put the flux densities on the \citet{baa77} scale.  This was done using (1) information contained within the reference, (2) the flux density of Cygnus A or Taurus A reported in the reference, or (3) the flux density of a bright source reported in the reference that had a flux density at a similar frequency from another reference for which the factor needed to put it on the \citet{baa77} scale was known.  Each reference with at least one match to a VLSS source is listed in Table \ref{reftab} with its \citet{baa77} scale factor, observing frequency, and CATS name.\par
As above, we searched the references in Table \ref{reftab} for matches to VLSS sources with a search radius of 40 arcsec.  Again, for each source with multiple components, the search was done separately for each component and the resulting spectra were combined into one composite spectrum.  For a source with matches from multiple references at the same observing frequency, the flux densities from the references were averaged and their uncertainties were added in quadrature.  From the list of references available through the CATS database, we excluded all references with observing frequencies of 327, 365, and 1400 MHz for all sources and those with an observing frequency of 4850 MHz for sources with declinations $<75^{\circ}$ (i.e., the area covered by the PMN and GB6 catalogs).  This was done to ensure that we would have reliable "anchor points" within the spectra at these frequencies to compensate for any spurious points obtained from smaller, possibly less complete surveys.\par
For each of the 388 sources with a peak brightness at 74 MHz $>$15 Jy beam$^{-1}$, the complete spectrum is given in Table \ref{spectab}.  For each source, we have also included an alternate name obtained from a search in the NASA/IPAC Extragalactic Database (NED), as well as the number of Gaussian components attributed to the sources, given as $N_{GC}$.  For each source, its spectrum is listed with each frequency given in MHz followed by a colon to separated these entries from the flux densities.  The flux densities and their 1$\sigma$ uncertainties as they were reported in their respective references are given in Jy.  Again, we emphasize that any source with a flux density listed at 327, 365, 1400, or 4850 MHz that is within the survey areas of the WENSS, TXS, NVSS, PMN, or GB6 catalogs, the flux density listed is taken from one of these five catalogs as described above.  Flux densities listed at other frequencies come from a multitude of references (see Table \ref{reftab}), and should be used with greater caution.  Flux densities at all frequencies are on the \citet{baa77} absolute flux density scale.  All spectra are also plotted in Fig. \ref{spec1}-\ref{spec16} with the name of each source at the top of each panel.  For sources with multiple components, the flux density at 74 MHz of the brightest component is also plotted separately as a $\times$ to provide an idea of how much the other components within the source contribute to the total flux density.

\subsection{Spectral Shapes}
To provide quantitative estimates of the shapes of the spectra presented here, we have used the following method to derive spectral parameters for the vast majority of the sources.  As a first step, for all sources with flux densities from the TXS or WENSS catalogs, the NVSS catalogs, and the PMN and/or GB6 catalogs, a power law was fit to these three higher frequency data points.  For sources at declinations larger than 75$^{\circ}$ which cannot be within the PMN or GB6 catalogs, a power law was fit to all data with $\nu>300$ MHz obtained from the CATS database searches.  The slopes of these fits, $\alpha_{>300}$, and the flux densities extrapolated to 74 MHz, $F_{ext}$, are listed for all sources, where applicable, in Table \ref{fittab}.  For sources without a relatively large number of available flux densities, the values of $\alpha_{>300}$ and the differences between $F_{ext}$ and the observed VLSS flux densities may serve as a way to roughly characterize the shapes of the spectra from lower to higher frequency.  The power-law fits are plotted with the data in the panels of Fig. \ref{spec1}-\ref{spec16} as dashed lines.\par
For sources with larger numbers of available flux densities, it is possible to determine a more detailed quantitative representation of their spectral shapes.  We have done this by fitting to the spectrum of each object with five or more flux densities (including the VLSS 74 MHz measurement) the following function
\begin{equation}
Y = A + BX + C \mbox{exp} \left (D X \right )
\end{equation}
where $Y=$log $F_{\nu}/$1 Jy and $X=$log $\nu/$74 MHz, using a Levenberg-Marquardt least-squares minimization routine.  This function was chosen rather than, for example, a higher order polynomial, because \citet{kue81} demonstrated that the spectra of the vast majority of bright radio sources are well approximated by either a simple power law or by equation (1).  Because of this, we will refer to spectral fits that use equation (1) as Kuehr fits for convenience.  For the case of spectra which "turn-over" at lower frequencies due to some form of spectral cutoff or absorption, the optical depth of the turnover at 74 MHz is given by $\tau=-C/$log $e$ and $\tau$ varies with frequency according to $\tau \propto \nu^{D \mbox{\scriptsize log }e}$.\par
For the Kuehr fits, we used the power-law fits to the $>300$ MHz data described above with the VLSS 74 MHz flux density to computed estimates for the $A$, $B$, and $C$ parameters.  The rms scatter about each of these power-law fits was also used with the uncertainty in the 74 MHz flux density to compute a 1$\sigma$ uncertainty in the $C$ parameter estimate.  The Kuehr fits were then performed using these initial guesses with an initial guess for $D$ set to -2.1/log $e$, as expected for the case of free-free absorption \citep{con92}.  The least-squares fitting routine used has the ability to perform the $\chi^{2}$ minimization while not allowing each of the free parameters to be outside of a given range of values.  We took advantage of this to ensure that Kuehr fits did not yield parameters that were in poor agreement with the $>300$ MHz data fits that used data from a shorter list of reliable references (i.e., Table \ref{mreftab}).  To achieve this, we used the following constraints on the free parameters.  For each case where $\alpha_{>300}<0$, $B$ was constrained to be between -10 and 0.5; for each case where $\alpha_{>300}\geq0$, $B$ was constrained to be between -0.5 and 10.  For sources where the estimate for $C$ plus its 1$\sigma$ uncertainty was $>0$, $C$ was constrained to be between -1 and 100.  For each case where the estimate for $C$ minus its 1$\sigma$ uncertainty was $<0$, $C$ was constrained to be between -100 and 1.  Finally, for each instance where the estimate for $C$ was within $\pm 1 \sigma$ of being zero, $C$ was constrained to be between -100 and 100.  For all fits, $D$ was constrained to be between -25 and 25.  All of these constraints were designed to ensure that the fitting routine worked properly and did not return parameters that were exceedingly unrealistic.\par
For each source, we also fit a simple power-law (i.e., $Y=A+BX$) to the entire spectrum to assess the need for the more complex model given in equation (1), using a standard linear least-squares fit.  For each of the Kuehr and power-law fits, the fit was performed twice.  The first fit was done in the standard way of weighting each flux density by $\sigma^{-2}$, where $\sigma$ is its estimated uncertainty.  However, in many cases, the estimated uncertainties in the flux densities given by the appropriate authors underestimate the amount of scatter in the actual spectra due to the effects of merging several catalogs of different sensitivities and spatial resolutions, the time variability of some sources, inherent complex spectral structures, and so on.  Because of this, the second fit was performed by weighting each flux density by its absolute deviation from the initial fit (in log-space) in order to give lower weight to possibly spurious data points.  For each Kuehr fit, this second fit was performed using the parameters from the first fit as initial guesses and using the same parameter constraints as were used in the first fit.\par
Since the estimated uncertainties in the flux densities were not used in the final fits, we have chosen not to assess the quality of the fits using the formal reduced $\chi^{2}$ of each fit.  Instead, we have chosen to use the following
\begin{equation}
\mbox{rms}_{Y-Y_{fit}} = \sqrt{\frac{\sum_{i=1}^{N_{\nu}} \left ( Y_{i} - Y_{fit,i} \right )^{2}}{N_{\nu}-N_{p}}}
\end{equation}
where $Y$ is the log of the observed flux density, $Y_{fit}$ is the fitted value at the same frequency, $N_{\nu}$ is the number of flux densities in the spectrum, and $N_{p}$ is the number of parameters in the fit (i.e., $N_{p}=$4 for Kuehr fits and $N_{p}=$2 for the power-law fits).  Given the inclusion of $N_{p}$ in the denominator of equation (2), this quantity is not strictly the root-mean-squared difference between the fit and the data, but since it is similar, we have elected to refer to this quantity as an "rms" for convenience.  For any source where rms$_{Y-Y_{fit}}$ is lower for the simple power-law fit than that for the Kuehr fit (which is true for about 56\% of the sources with $N_{\nu}>5$), we have dropped the values for $C$ and $D$ and have adopted values for $A$ and $B$ from the power-law fit.  The resulting values for $A$, $B$, $C$, and $D$ are listed in Table \ref{fittab} along with values of rms$_{Y-Y_{fit}}$.  The fitted curves corresponding to these parameters are plotted in the panels of Fig. \ref{spec1}-\ref{spec16} as red solid lines.  For cases where rms$_{Y-Y_{fit}}>$log 1.5 (i.e., the fit and the data differ by more than 50\% on average), 31 sources in all, the curves are blue.

\section{Discussion}
\begin{figure}
\plotone{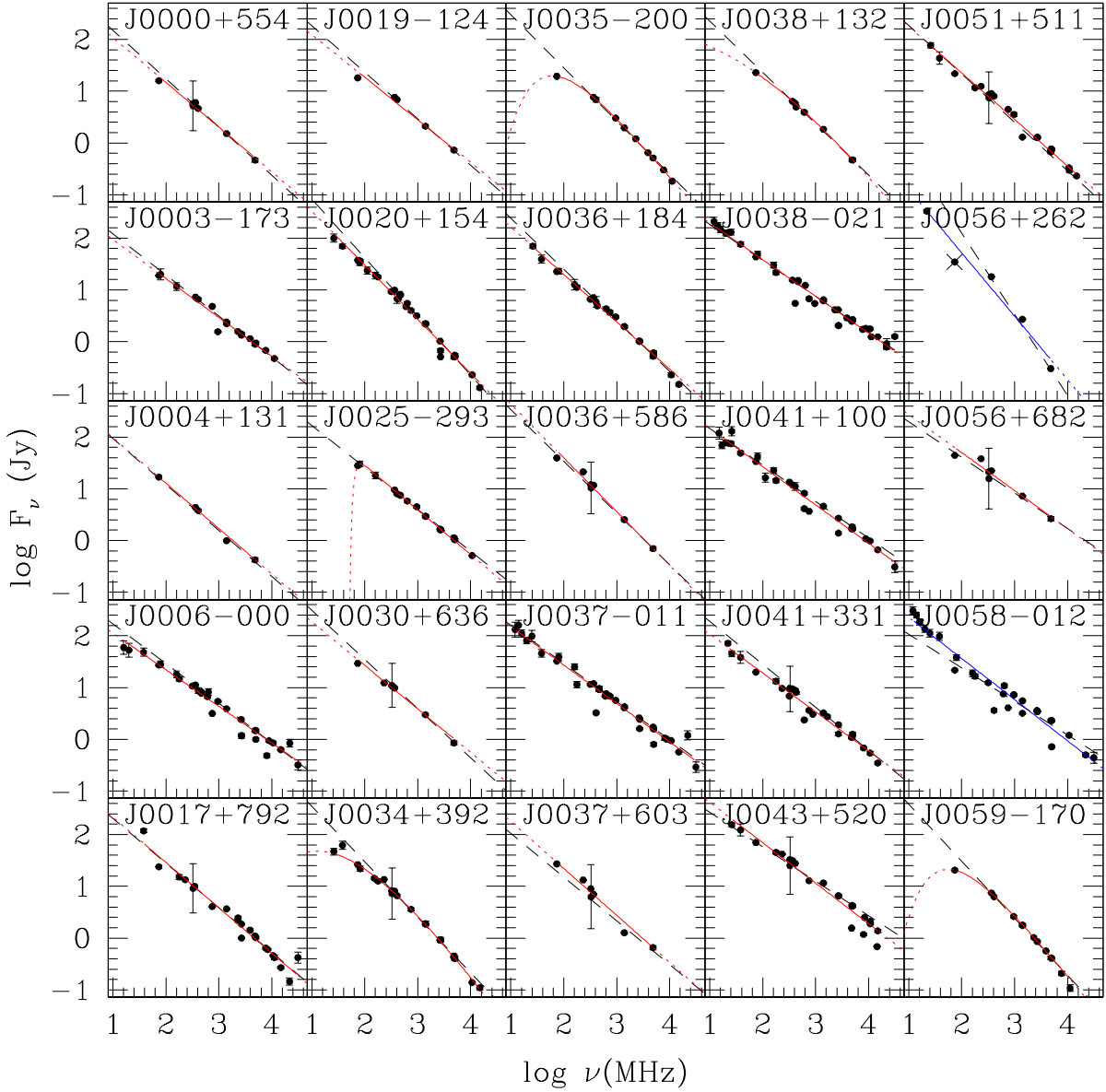}
\caption{Spectra for the first 25 sources given in Table \ref{spectab}.  For each source with more than one Gaussian component, the 74 MHz flux density of the brightest component is plotted as a $\times$.  For sources with data in the TXS or WENSS catalogs, the NVSS catalog, and the PMN or GB6 catalogs (see test and Table \ref{mreftab}), a linear fit to these data is plotted as a dashed line.  For sources with eight or more flux density measurements, the best fitting curve (see text) is plotted as a red line.  The dotted portions of these curves indicate the regions where there is no data and where the fitted curves should be used with extreme caution (if at all).  For sources where the rms difference between the data and the fit is $>$log 1.5 ($\approx 0.18$; 31 sources in all), the best fitting curve is plotted in blue.  The parameters of these fits are listed in Table \ref{fittab}.}
\label{spec1}
\end{figure}

\begin{figure}
\plotone{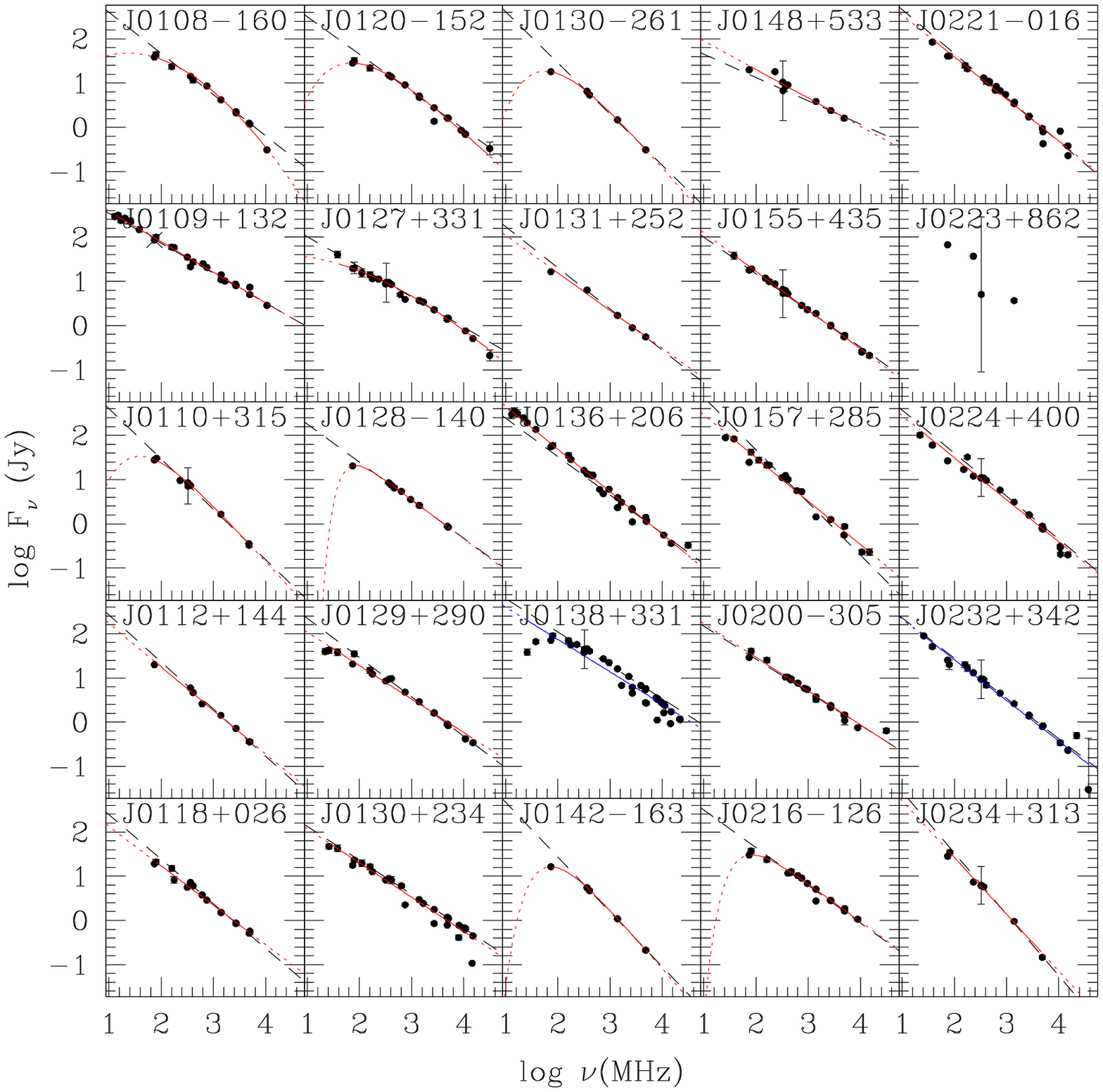}
\caption{The same as Fig. \ref{spec1}, but for the next 25 sources.}
\label{spec2}
\end{figure}

\begin{figure}
\plotone{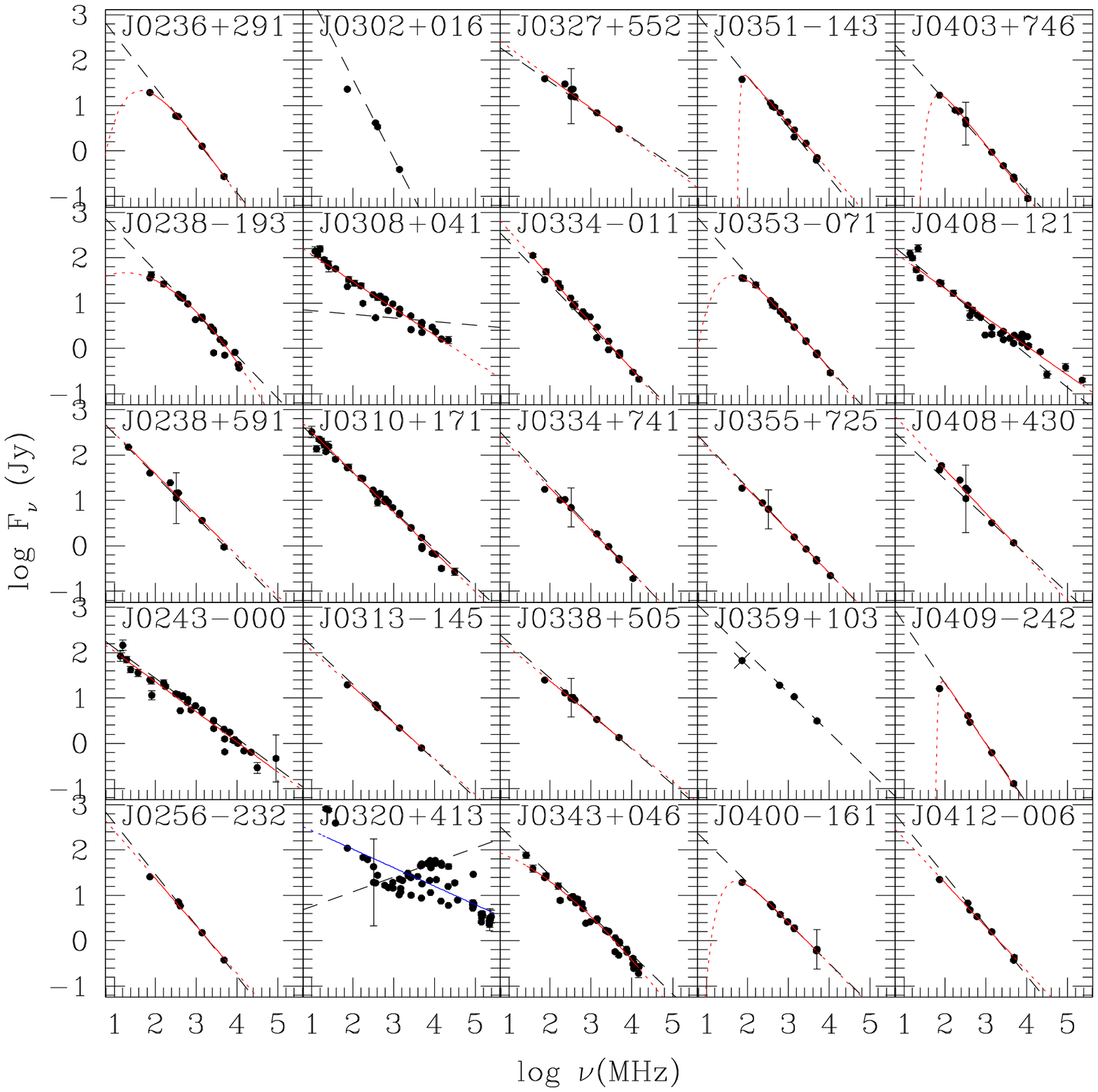}
\caption{The same as Fig. \ref{spec1}, but for the next 25 sources.}
\label{spec3}
\end{figure}

\begin{figure}
\plotone{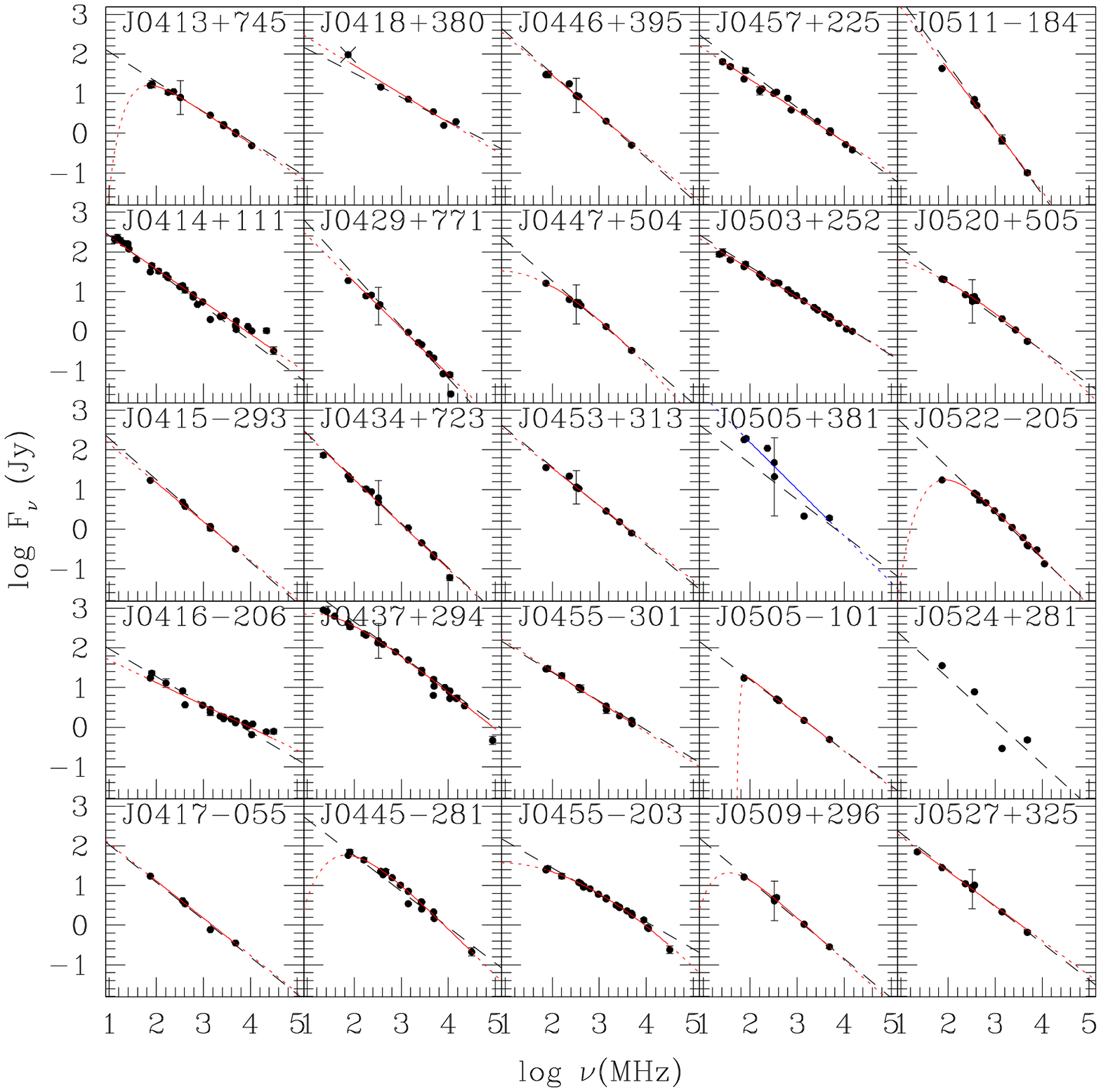}
\caption{The same as Fig. \ref{spec1}, but for the next 25 sources.}
\label{spec4}
\end{figure}

\begin{figure}
\plotone{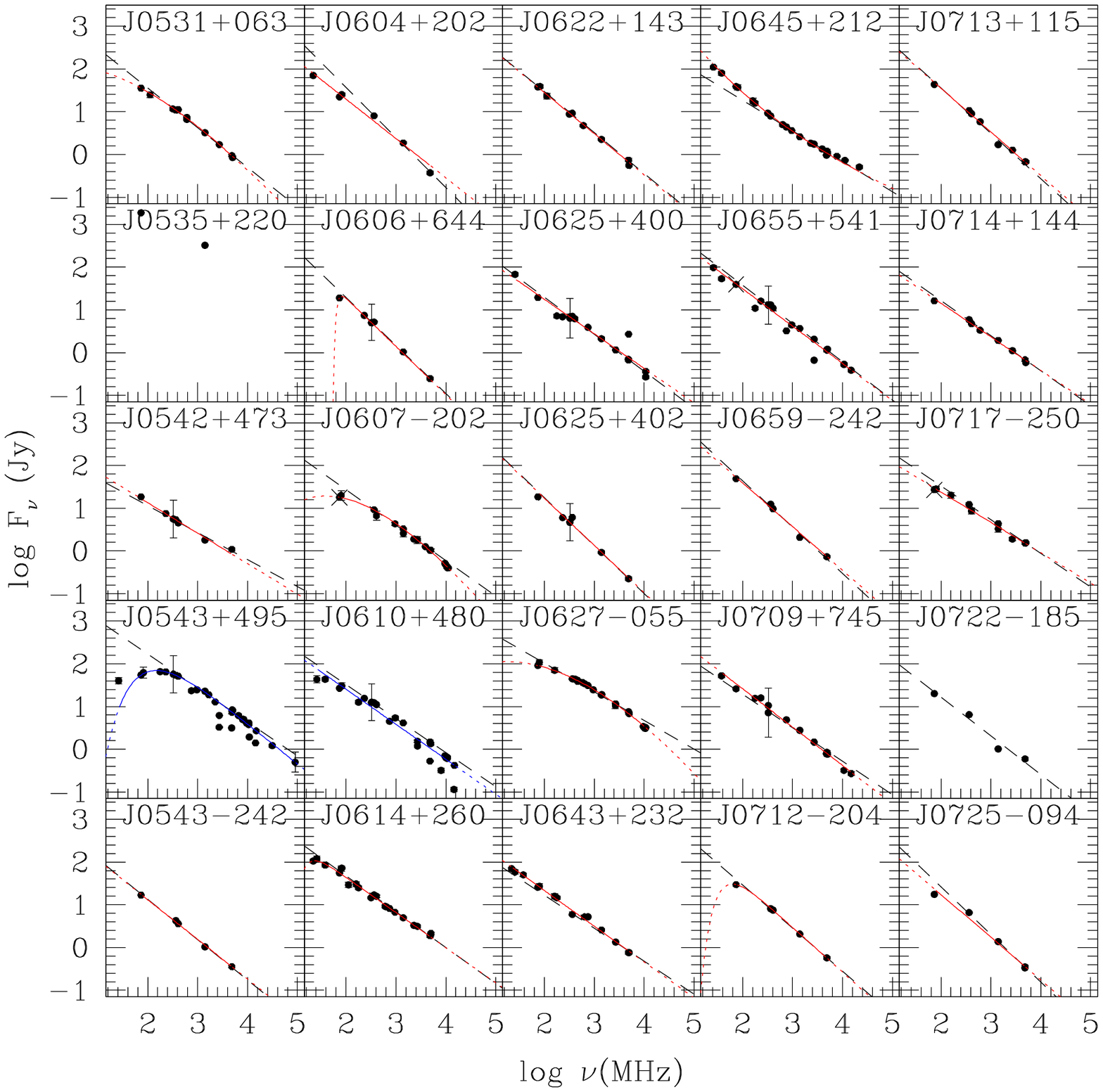}
\caption{The same as Fig. \ref{spec1}, but for the next 25 sources.}
\label{spec5}
\end{figure}

\begin{figure}
\plotone{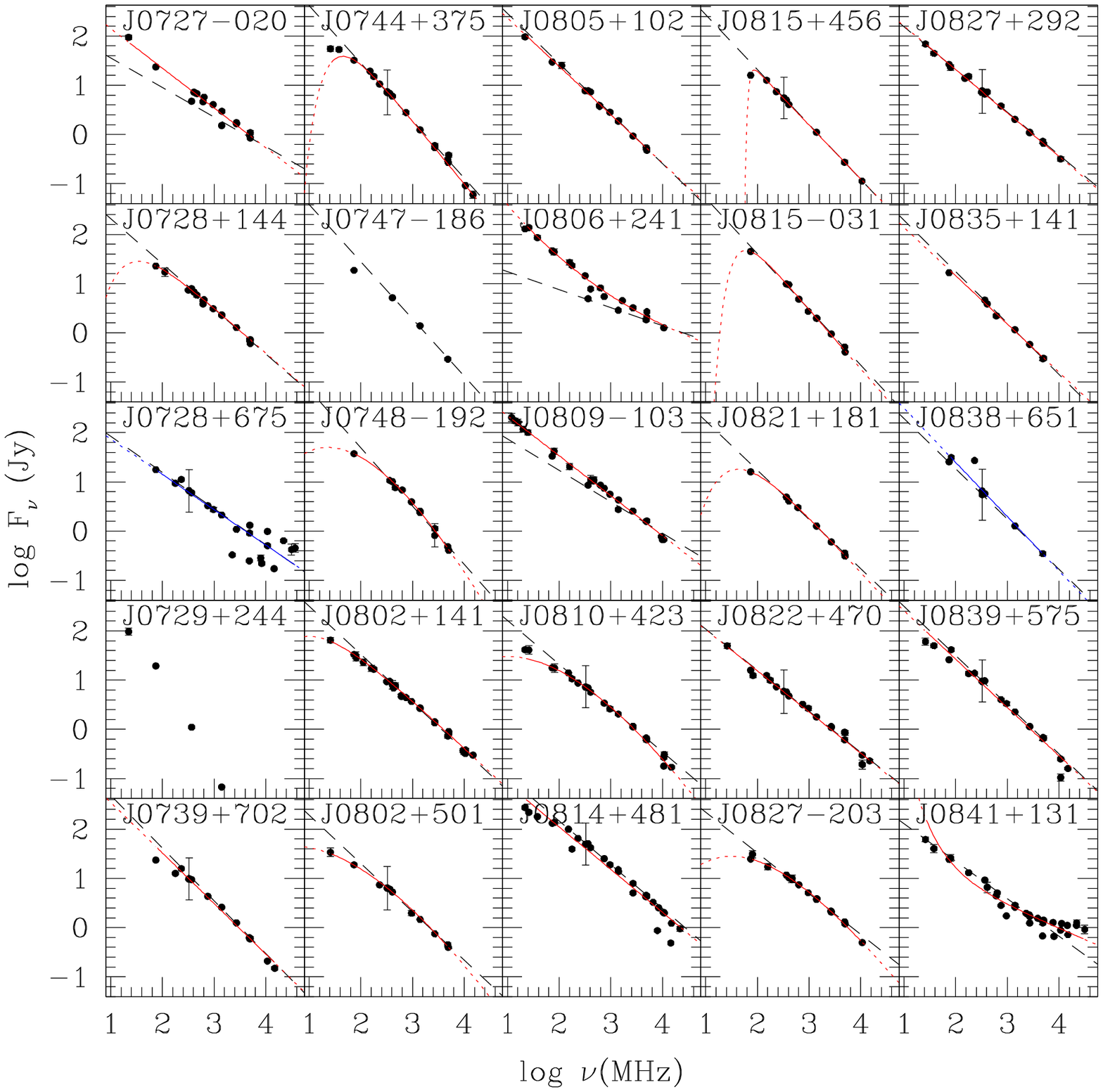}
\caption{The same as Fig. \ref{spec1}, but for the next 25 sources.}
\label{spec6}
\end{figure}

\begin{figure}
\plotone{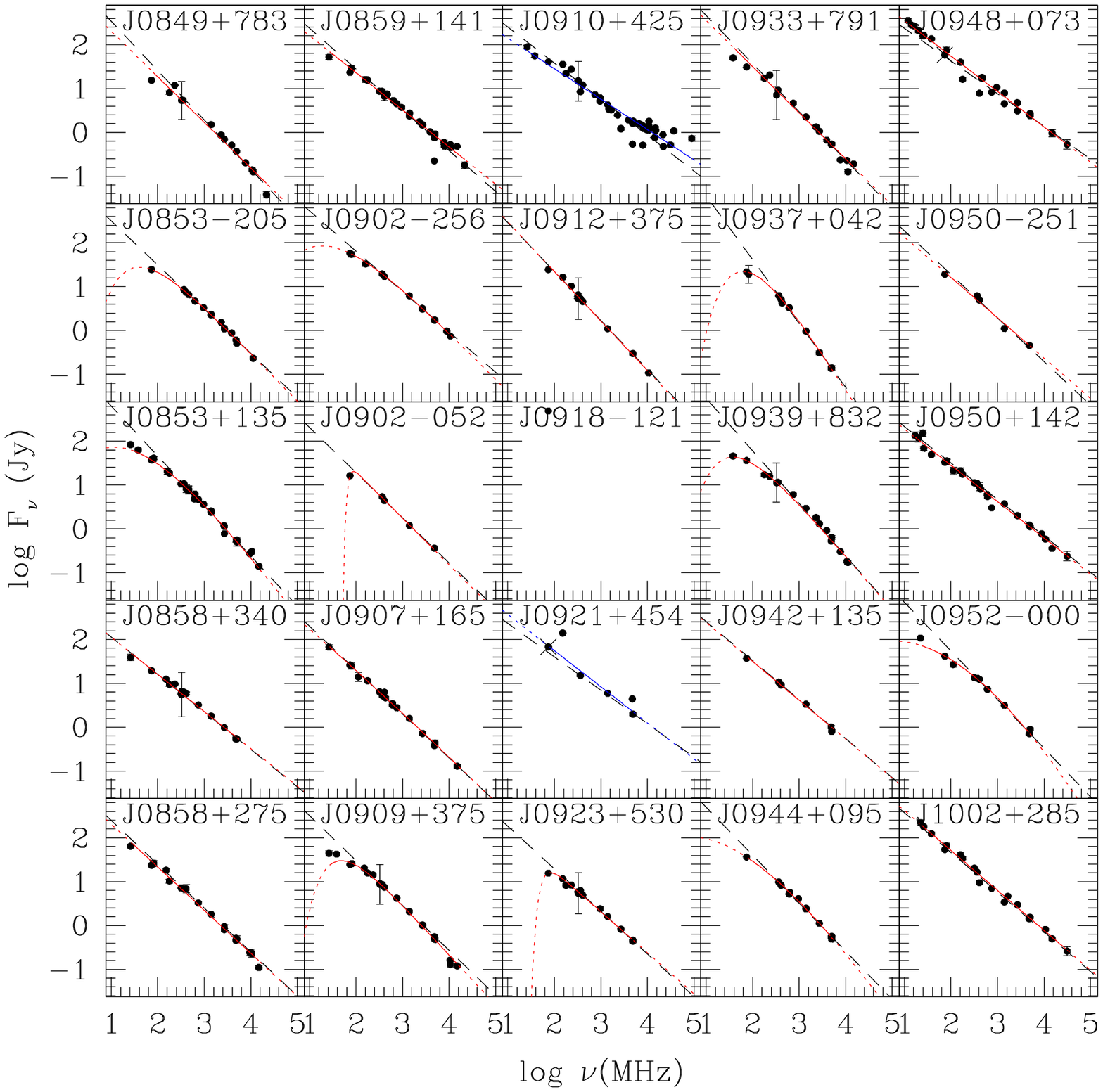}
\caption{The same as Fig. \ref{spec1}, but for the next 25 sources.}
\label{spec7}
\end{figure}

\begin{figure}
\plotone{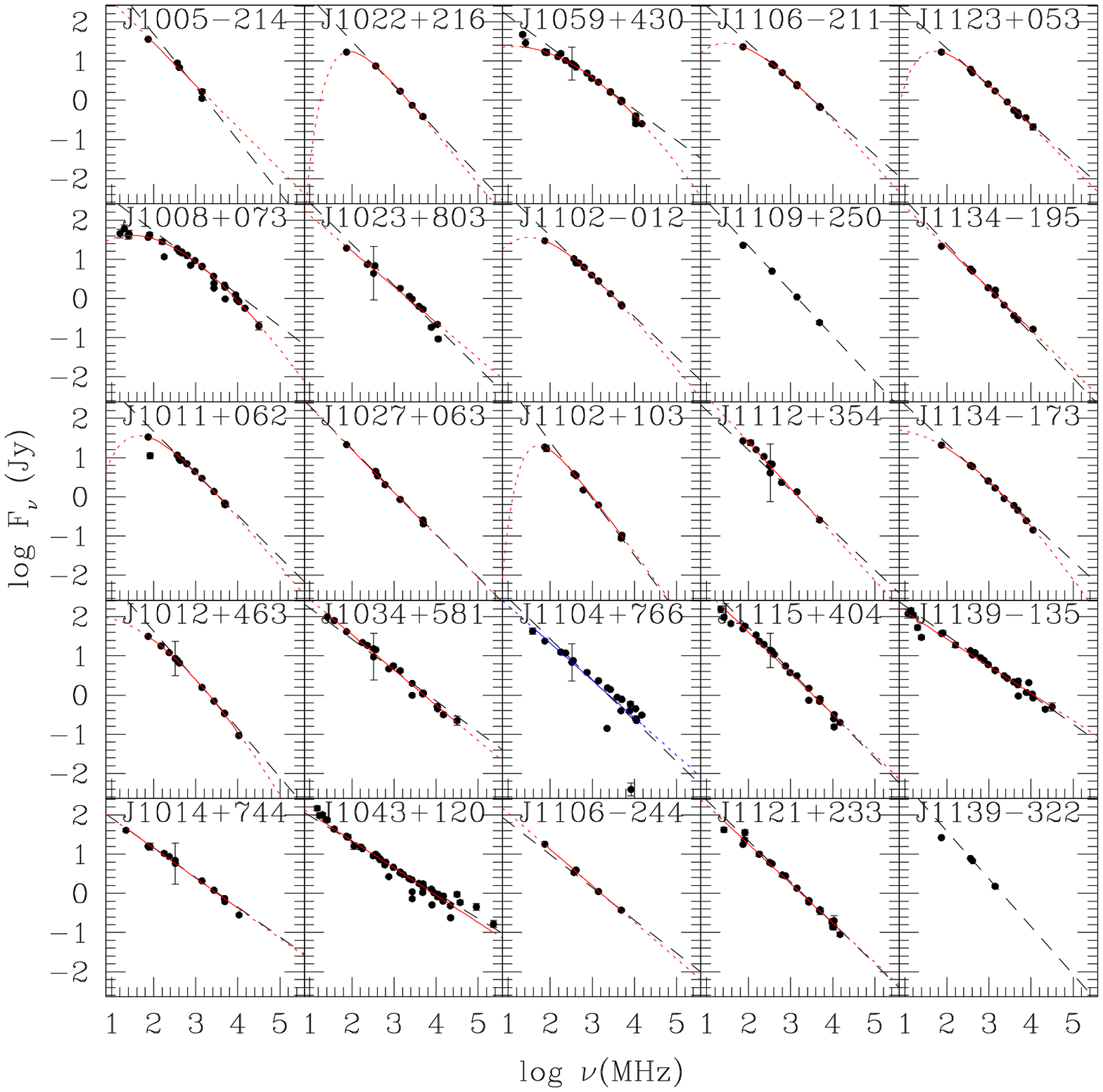}
\caption{The same as Fig. \ref{spec1}, but for the next 25 sources.}
\label{spec8}
\end{figure}

\begin{figure}
\plotone{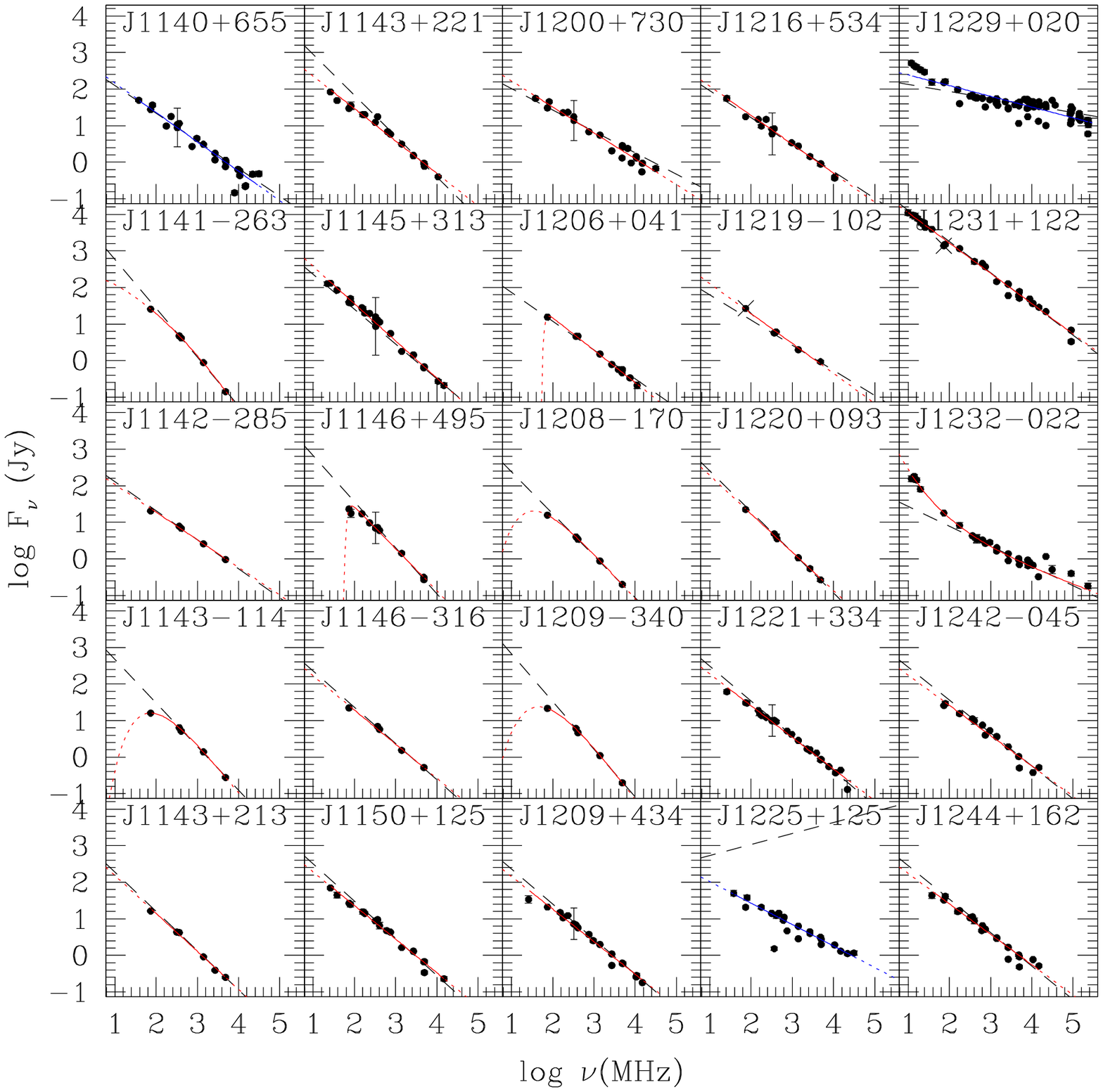}
\caption{The same as Fig. \ref{spec1}, but for the next 25 sources.}
\label{spec9}
\end{figure}

\begin{figure}
\plotone{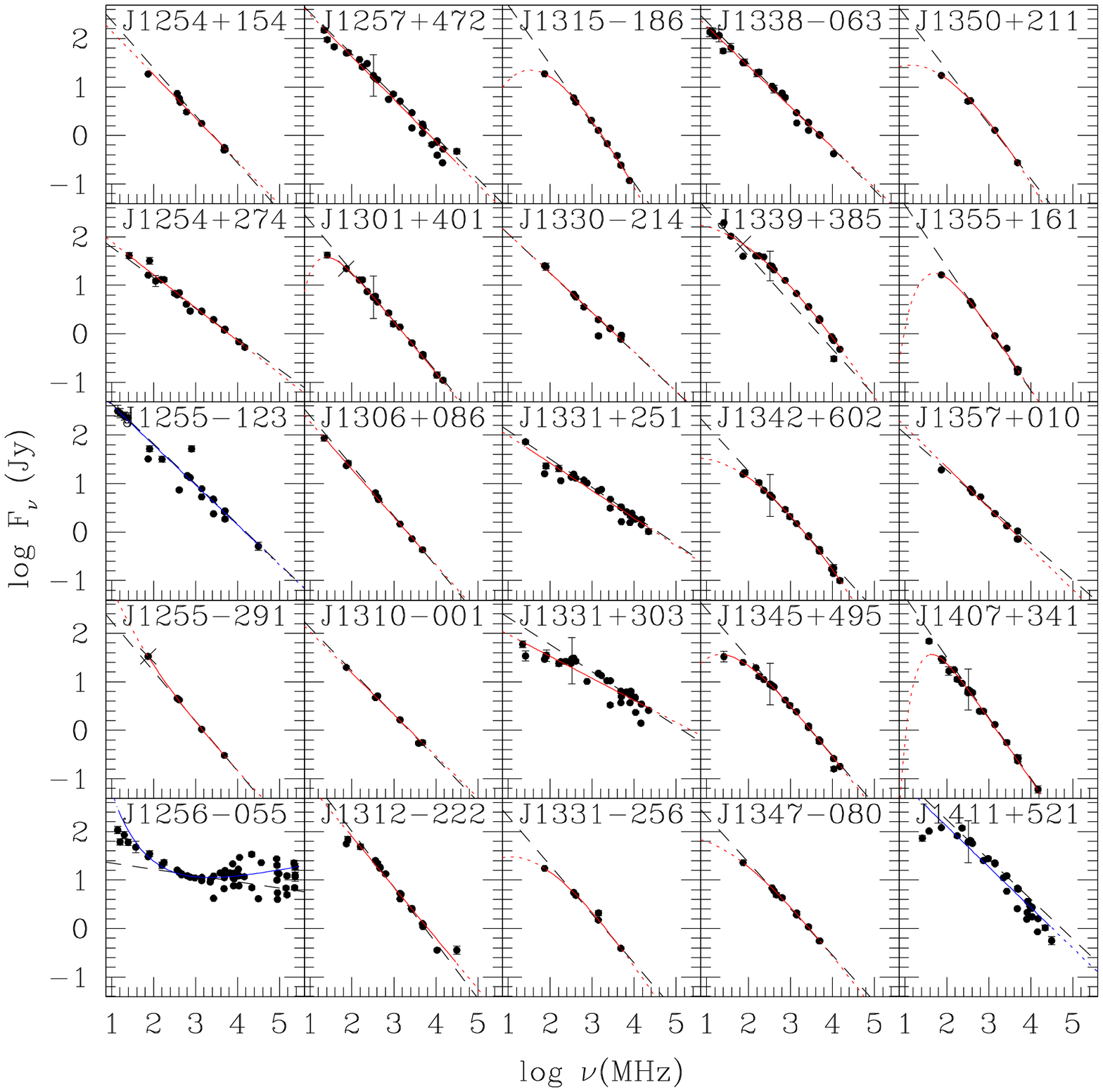}
\caption{The same as Fig. \ref{spec1}, but for the next 25 sources.}
\label{spec10}
\end{figure}

\begin{figure}
\plotone{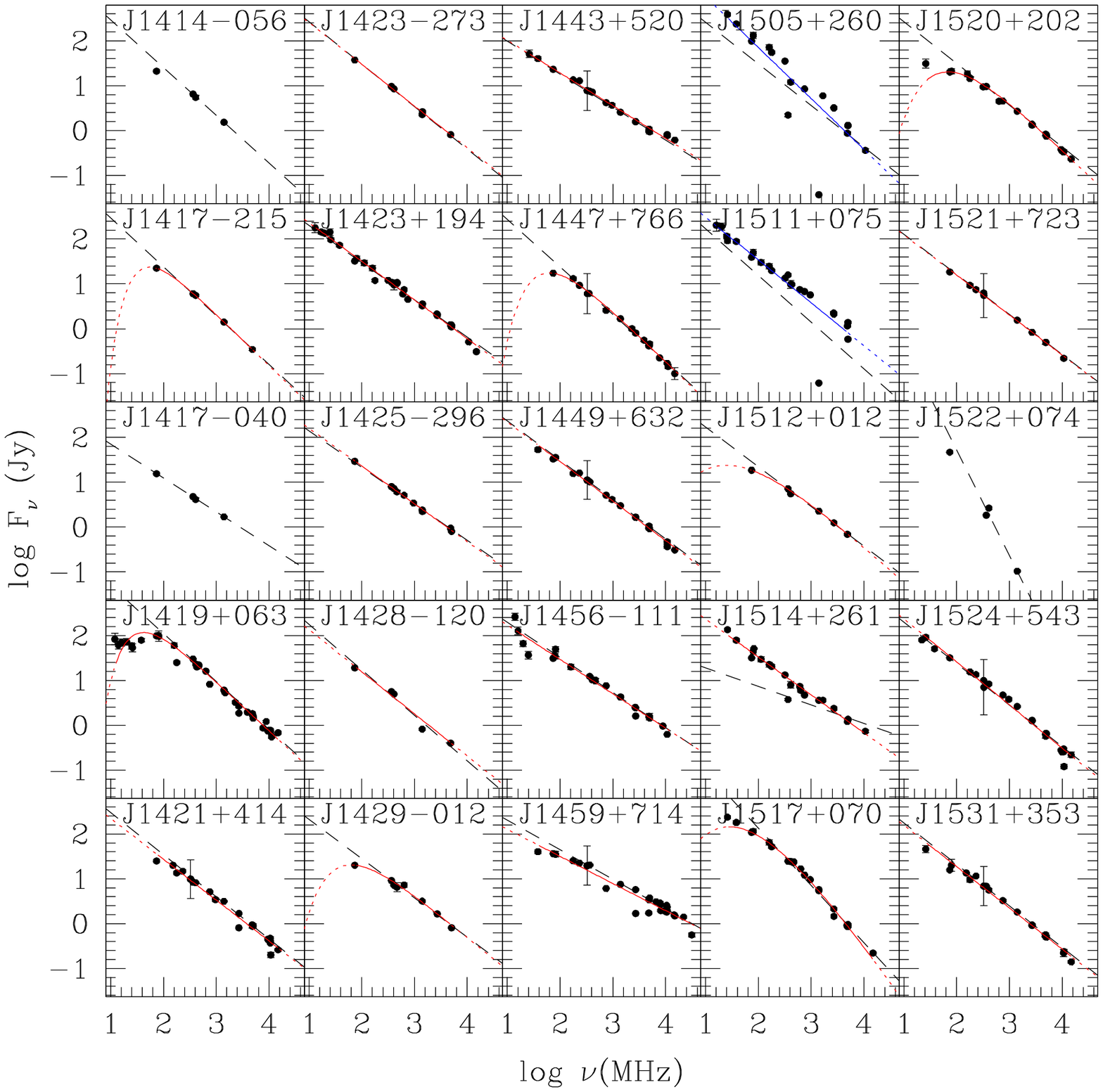}
\caption{The same as Fig. \ref{spec1}, but for the next 25 sources.}
\label{spec11}
\end{figure}

\begin{figure}
\plotone{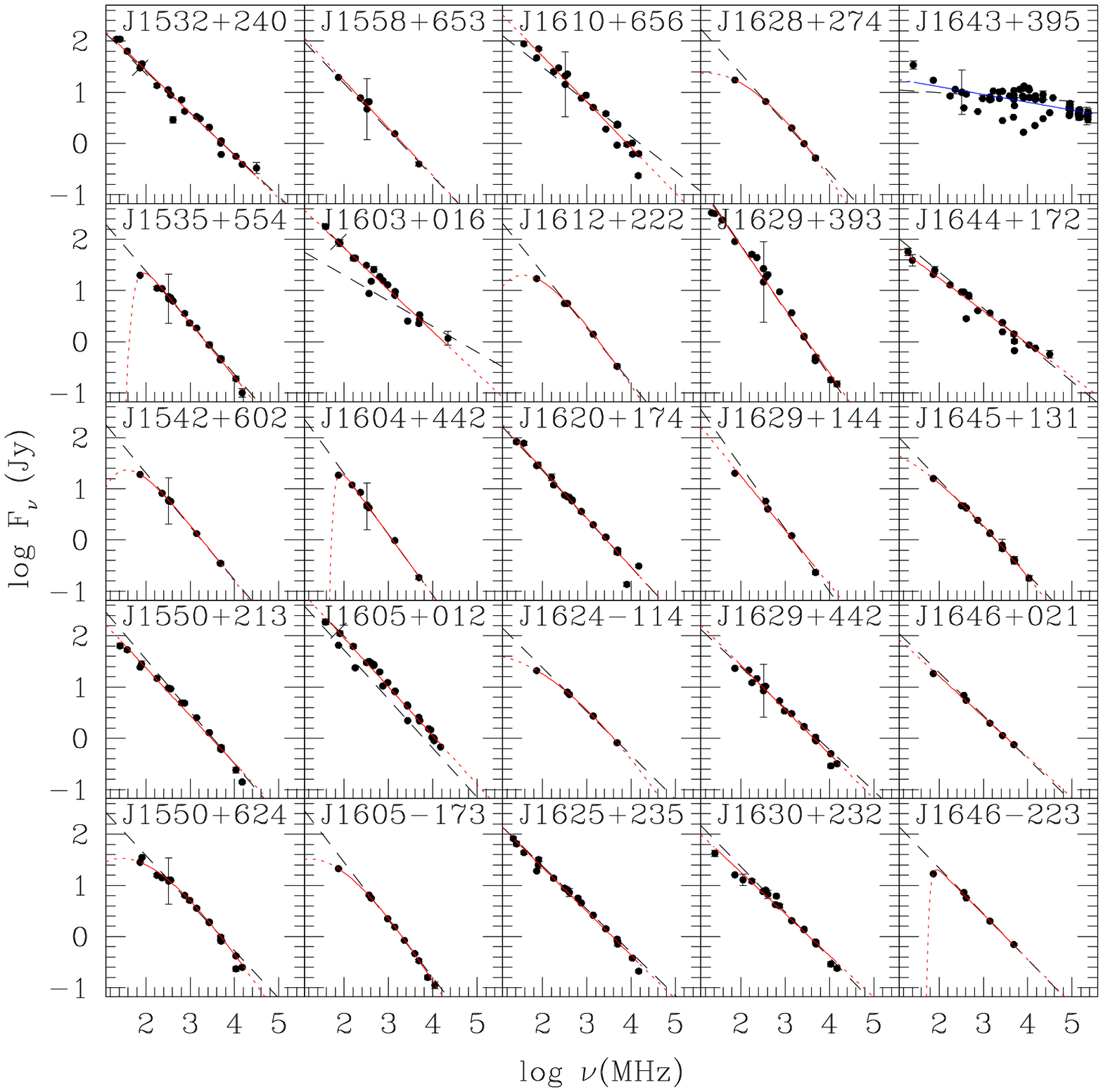}
\caption{The same as Fig. \ref{spec1}, but for the next 25 sources.}
\label{spec12}
\end{figure}

\begin{figure}
\plotone{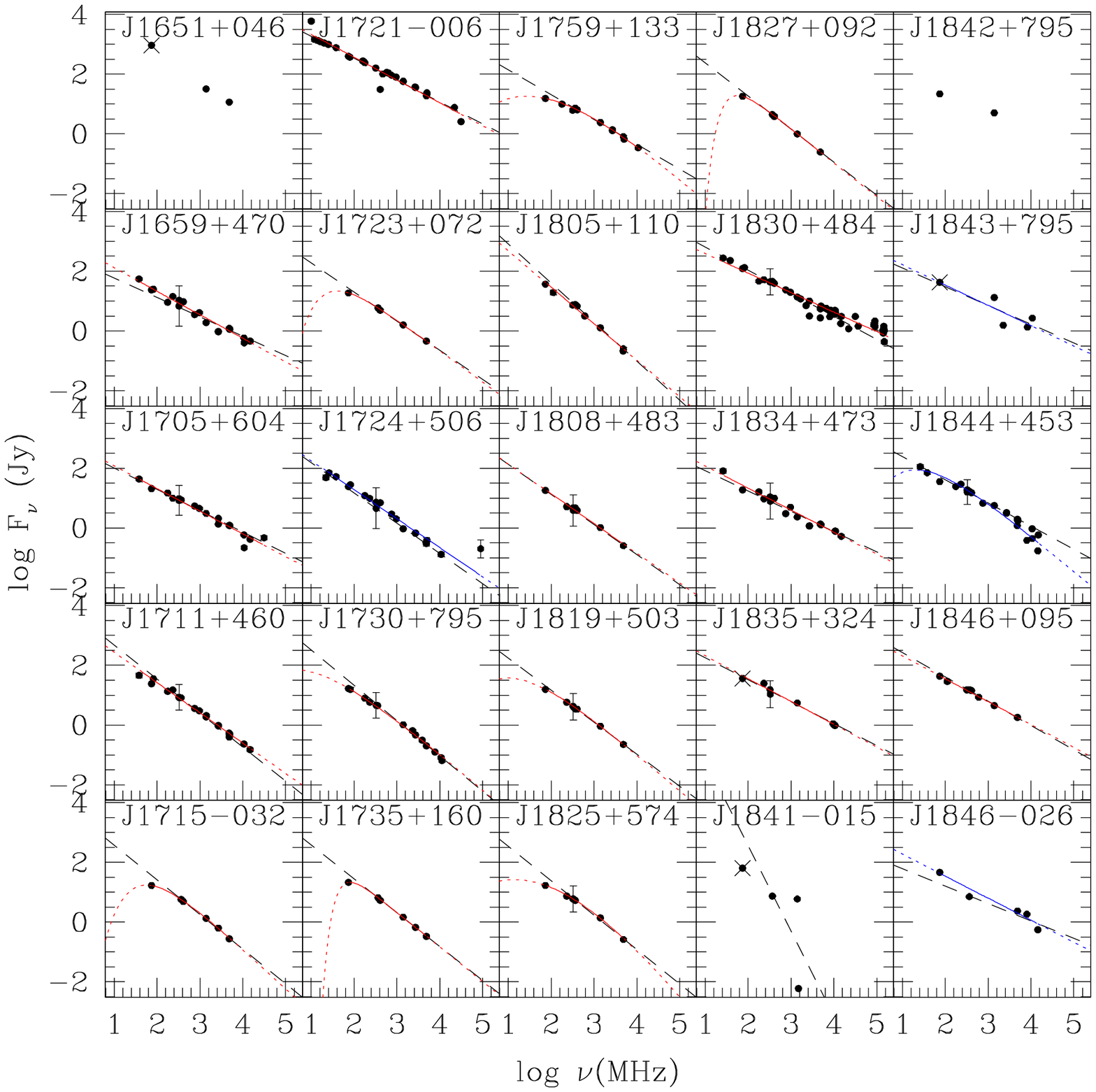}
\caption{The same as Fig. \ref{spec1}, but for the next 25 sources.}
\label{spec13}
\end{figure}

\begin{figure}
\plotone{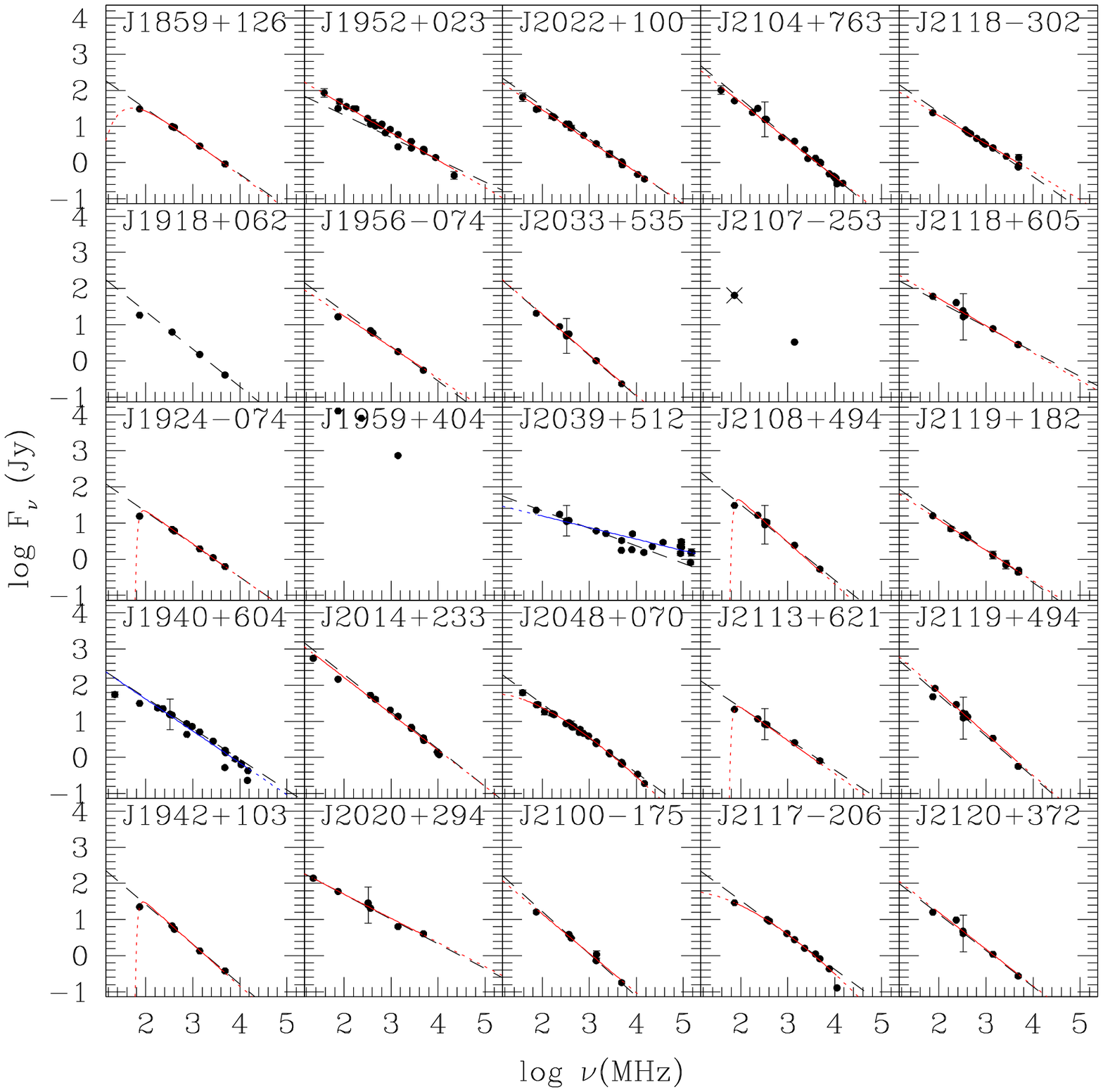}
\caption{The same as Fig. \ref{spec1}, but for the next 25 sources.}
\label{spec14}
\end{figure}

\begin{figure}
\plotone{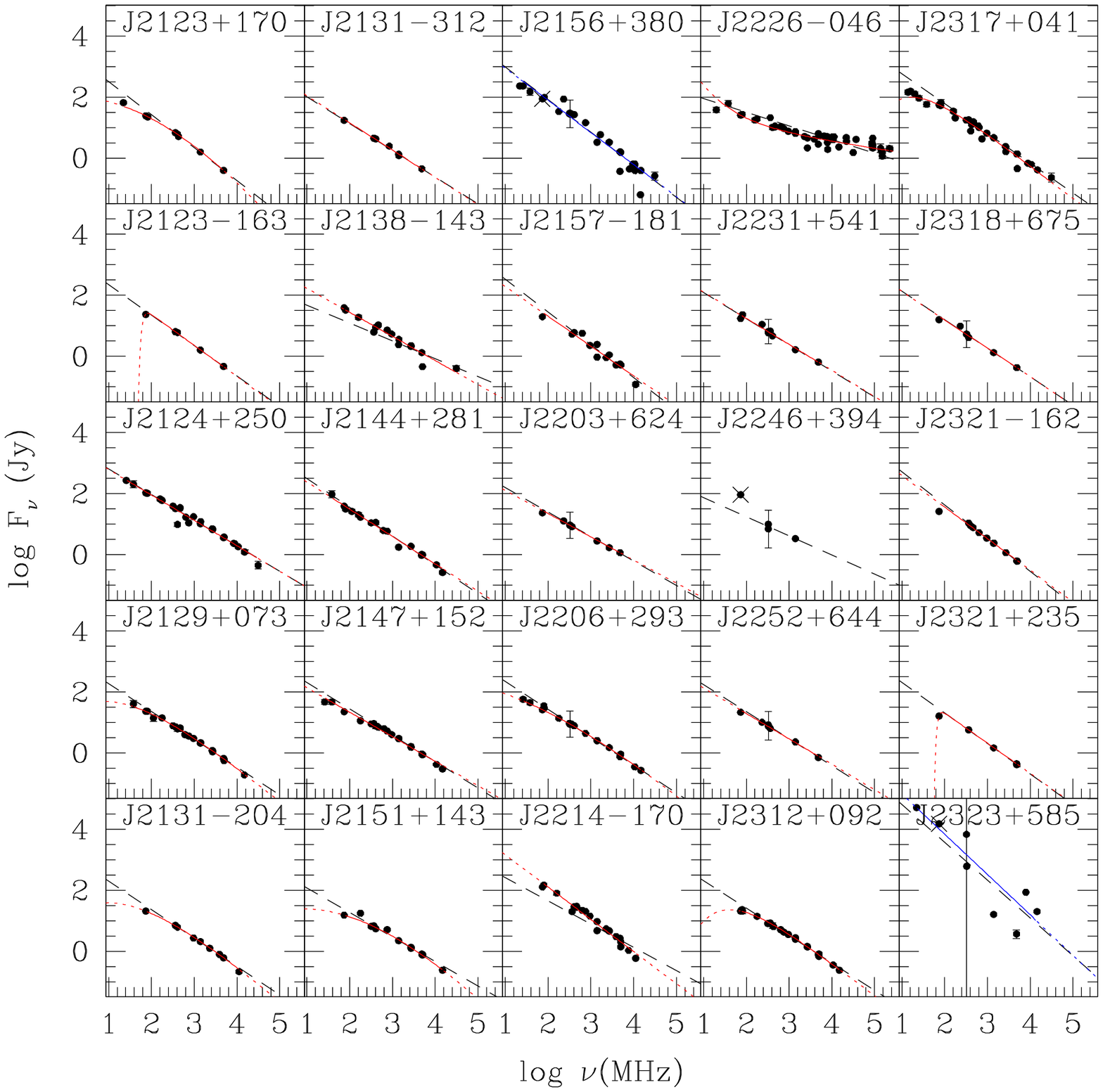}
\caption{The same as Fig. \ref{spec1}, but for the next 25 sources.}
\label{spec15}
\end{figure}

\begin{figure}
\plotone{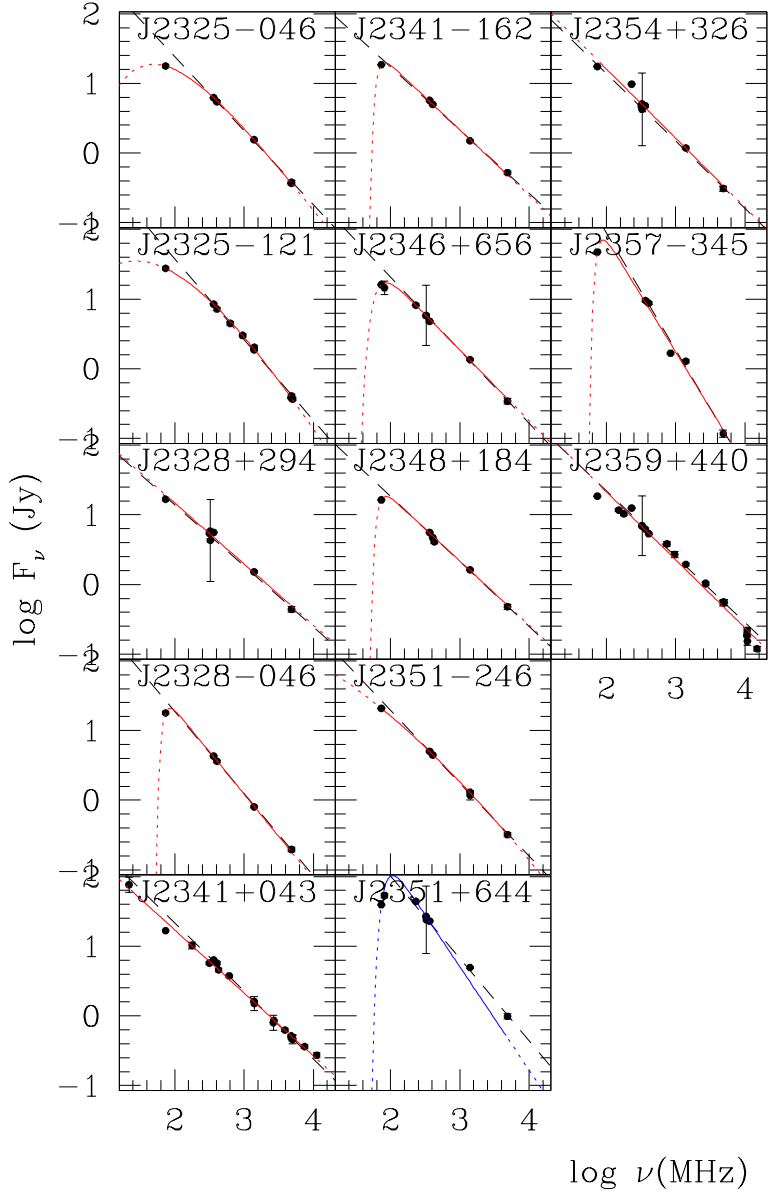}
\caption{The same as Fig. \ref{spec1}, but for the final 13 sources.}
\label{spec16}
\end{figure}

The spectra for the 388 bright 74 MHz sources presented here are provided as a service to the low frequency radio astronomy community.  Because these spectra have improved accuracy at low
frequencies, they will provide and excellent list of possible calibration sources for current (e.g., GMRT, VLA) and future (e.g., EVLA, LWA, LOFAR, MWA) facilities capable of observing at frequencies $^{<}_{\sim}100$ MHz.  In general, the spectral fits discussed above will provide an adequate means of interpolating to frequencies not specifically contained in the spectra as they are presented here.  However, we caution the reader that using these fits to extrapolate beyond the frequency ranges of the spectra used to generate the fits is not advisable.  This can be seen in several examples in the panels of Fig. \ref{spec1}-\ref{spec16} where the fitted curve turns dramatically upward or downward beyond the lowest or highest frequency data point.  One must also be wary of sources where rms$_{Y-Y_{fit}}$ is abnormally high.  In Fig. \ref{spec1}-\ref{spec16}, we have highlighted the 31 cases where rms$_{Y-Y_{fit}}>$log 1.5.  In many of these cases, one or two aberrant points have inflated the value of rms$_{Y-Y_{fit}}$.  However, there are other cases where the shapes of the spectra were too complex to fit with equation (1).  Possibly the best example of this is VLSS J031948.1+413042.5 (see Fig. \ref{spec3}, bottom row, second from the left).  The complicated "S" shape of this source's spectrum was fit so poorly by equation (1) that formally, a simple power-law fit the spectrum better.  In these instances, the reader is encouraged to use an alternative method of interpolation.  We also stress that we have included the rms$_{Y-Y_{fit}}$ values in Table \ref{fittab} so that one may use different limits on the quality of the fits (i.e., instead of rms$_{Y-Y_{fit}}>$log 1.5) to determine which spectra have fitted spectral parameters that can be trusted. \par
Finally, we acknowledge that the spectral parameters presented here also allow for a more detailed study of the properties and physical nature of low frequency radio sources.  For instance, there are some sources which have virtually no higher frequency data, implying very steep-spectrum sources.  Many of the spectra have shapes that change from lower to higher frequencies, implying that processes such as intrinsic or extrinsic free-free absorption, synchrotron self-absorption, and spectral aging may play a role within these objects.  However, we have set aside such analyses to subsequent papers where the spectral properties may be discussed in more detail and, when completed, the entire VLSS catalog may be used.

\acknowledgements  
The authors would like to thank K. Kellerman and W. Erickson for useful comments and suggestions.  This research was performed while the lead author held a National Research Council Research Associateship Award at the Naval Research Laboratory.  Basic research in astronomy at the Naval Research Laboratory is supported by 6.1 base funding.  The National Radio Astronomy Observatory is a facility of the National Science Foundation operated under cooperative agreement by Associated Universities, Inc.  This research has made use of the NASA/IPAC Extragalactic Database (NED) which is operated by the Jet Propulsion Laboratory, California Institute of Technology, under contract with the National Aeronautics and Space Administration.

\clearpage
\thispagestyle{empty}
\begin{deluxetable}{l@{$\;\;$}l@{$\;\;$}c@{$\;\;$}c@{$\;\;$}c@{$\;\;$}c@{$\;\;$}c@{$\;\;$}l}
\rotate
\tabletypesize{\footnotesize}
\tablecolumns{8}
\tablewidth{0pc}
\tablecaption{High Frequency References}
\tablehead{
\colhead{} & \colhead{} & \colhead{$\nu$} & \colhead{} & \colhead{FWHM$_{\mbox{\scriptsize B}}$} & \colhead{$\overline{\mbox{rms}}$} & \colhead{} & \colhead{} \\
\colhead{Name} & \colhead{Abbrev.} & \colhead{(GHz)} & \colhead{$f_{Baars}$} & \colhead{(arcsec)} & \colhead{(mJy/beam)} & \colhead{Dec. Range} & \colhead{Reference} \\
\colhead{(1)} & \colhead{(2)} & \colhead{(3)} & \colhead{(4)} & \colhead{(5)} & \colhead{(6)} & \colhead{(7)} & \colhead{(8)}}
\startdata
The Westerbork Northern Sky Survey & WENSS &   0.325 & 1 & 54 & 3.6 & $\delta>30^{\circ}$ & \citet{ren97} \\
The Texas Survey of Radio Sources & TXS &   0.365 & 1.041 & $\sim 60$ & 20 & $-35.5^{\circ}<\delta<71.5^{\circ}$ & \citet{dou96} \\
The NRAO VLA Sky Survey & NVSS & 1.4 & 1 & 45 & 0.45 & $\delta>-40^{\circ}$ & \citet{con98} \\
The Parkes-MIT-NRAO Sky Surveys & PMN & 4.85 & 1 & 250 & $\sim$7 & $-87.5^{\circ}<\delta<-37^{\circ}$ & \citet{wri94} (PMN S) \\
 &    &   4.85 & 1 & 250 & 14 & $-37^{\circ}<\delta<-29^{\circ}$ & \citet{wri96} (PMN Z) \\
 &    &   4.85 & 1 & 250 & 8 & $-29^{\circ}<\delta<-9.5^{\circ}$ & \citet{wri94} (PMN T) \\
 &    &   4.85 & 1 & 250 & 8 & $-9.5^{\circ}<\delta<10^{\circ}$ & \citet{gri95} (PMN E) \\
\enddata
\tablecomments{Col. (1): Full name of the survey used.  Col. (2): Abbreviated survey name.  Col. (3): Survey observing frequency, in GHz.  Col. (4): Multiplicative factor applied to the flux densities to put them on the Baars et al.\ (1977) flux density scale.  Col. (5): Approximate full width and half maximum (FWHM) of the survey beam.  Col. (6): Approximated typical rms noise for the survey observations/images.  Col. (6): The declination range of the survey.  Col. (8): The reference for the survey.}
\label{mreftab}
\end{deluxetable}

\clearpage
\begin{deluxetable}{l@{$\;\;$}c@{$\;\;$}c@{$\;\;$}l@{$\;\;$}}
\tabletypesize{\small}
\tablecolumns{4}
\tablewidth{0pc}
\tablecaption{References for Spectral Data}
\tablehead{
\colhead{} & \colhead{} & \colhead{$\nu$} & \colhead{} \\
\colhead{Name} & \colhead{$f_{Baars}$} & \colhead{(MHz)} & \colhead{Reference} \\
\colhead{(1)} & \colhead{(2)} & \colhead{(3)} & \colhead{(4)}}
\startdata
\sidehead{Single Frequency Catalogs}
\hline
                 UTR &                   1 &                                              16.7 &  Braude, et al.\   1978, 85, Ap\&SS,  \\
               RCS86 &               0.992 &                                                22 &  Roger, et al.\   1986, A\&AS, 65, 485 \\
                  VE &                   1 &                                              26.3 &  Viner \& Erickson  1975, AJ, 80, 931 \\
                 NEK &                   1 &                                              30.9 &  Kassim  1988, ApJS, 68, 715 \\
                  8C &               0.818 &                                                38 &  Hales, et al.\  1995, MNRAS, 274, 447 \\
               IPS81 &                   1 &                                              81.5 &  Purvis, et al.\   1987, MNRAS, 229, 589 \\
                RH74 &                   1 &                                              81.5 &  Readhead  1974, MmRAS, 78, 1 \\
                6CII &               0.904 &                                               151 &  Hales, et al.\   1988, MNRAS, 234, 919 \\
\enddata

\tablecomments{Col. (1): Name of the catalog used in the CATS database.  Col. (2): Multiplicative factor applied to the flux densities to put them on the Baars et al.\ (1977) flux density scale.  Col. (3): The observing frequncy/ies of the catalog.  Col. (4): The reference for the catalog.}
\label{reftab}
\end{deluxetable}

\begin{deluxetable}{r@{$\;\;$}r@{$\;\;$}r@{$\;\;$}r@{$\;\;$}r@{$\;\;$}r@{$\;\;$}r@{$\;\;$}r@{$\;\;$}r@{$\;\;$}r@{$\;\;$}r@{$\;\;$}r@{$\;\;$}r@{$\;\;$}r@{$\;\;$}r@{$\;\;$}r@{$\;\;$}r@{$\;\;$}r@{$\;\;$}r@{$\;\;$}r@{$\;\;$}r@{$\;\;$}r@{$\;\;$}r}
\tabletypesize{\tiny}
\tablecolumns{23}
\tablewidth{0pc}
\tablecaption{The Spectra}
\tablehead{}
\startdata
\sidehead{J0000+554$\;\;\;\;$   87GB 235746.2+552220  , $N_{GC}=$1, $\alpha=$00:00:20.4 $\delta=$+55:39:06.1}
     74: &                15.92 &                 0.37 & &     327: &                 5.20 &                 5.78 & &     408: &                 4.61 &                 0.03 & &    4850: &                 0.47 &                 0.04 & &         &                      &                      & &         &                      &                     \\
    325: &                 5.78 &                 0.01 & &     365: &                 6.03 &                 0.14 & &    1400: &                 1.52 &                 0.05 & &         &                      &                      & &         &                      &                      & &         &                      &                     \\
\\
\sidehead{J0003-173$\;\;\;\;$  *[HB89] 0000-177       , $N_{GC}=$1, $\alpha=$00:03:22.1 $\delta=$-17:27:11.1}
     74: &                18.98 &                 0.11 & &     365: &                 7.27 &                 0.14 & &     960: &                 1.57 &                 0.06 & &    2300: &                 1.58 &                 0.06 & &    3900: &                 1.15 &                 0.02 & &    7700: &                 0.68 &                 0.02\\
     80: &                20.00 &                 5.00 & &     408: &                 6.59 &                 0.27 & &    1400: &                 2.41 &                 0.07 & &    2695: &                 1.34 &                 0.04 & &    4850: &                 0.89 &                 0.05 & &   11200: &                 0.47 &                 0.01\\
    160: &                11.70 &                 2.00 & &     750: &                 4.83 &                 0.18 & &    1410: &                 2.23 &                 0.11 & &    2700: &                 1.40 &                 0.10 & &    5009: &                 0.95 &                 0.07 & &         &                      &                     \\
\\
\enddata
\tablecomments{For each of the 388 VLSS sources, the spectral data are listed here preceeded by the VLSS source name, an alternate name taken from NED, the number of Gaussian components from the VLSS image attributed to the source, and the J2000 right ascension, $\alpha$, in hours, minutes, and seconds and declination, $\delta$, in degrees, arcminutes, and arcseconds.  For each flux density, the frequency is listed, followed by a colon, with the flux density and the 1$\sigma$ uncertainty in that flux density, listed directly to the right in units of Jy.  All flux densities were obtained from the CATS database with the exception of flux densities at 327, 365, 1400, and 4850 MHz which were taken from the TXS, WENSS, NVSS, and GB6 or PMN catalogs where applicable (see text).  See Table \ref{reftab} for a list of the references used in the CATS database searches.}
\label{spectab}
\end{deluxetable}
\begin{deluxetable}{lccccccccc}
\tabletypesize{\footnotesize}
\tablecolumns{10}
\tablewidth{0pc}
\tablecaption{Parameters of the Spectral Fits}
\tablehead{\colhead{} & \colhead{} & \colhead{$F_{74}$} & \colhead{} & \colhead{$F_{ext}$} & \colhead{} & \colhead{} & \colhead{} & \colhead{} & \colhead{rms$_{Y-Y_{fit}}$}\\
\colhead{Name} & \colhead{$N_{\nu}$} & \colhead{(Jy)} & \colhead{$\alpha_{>300}$} & \colhead{(Jy)} & \colhead{A} & \colhead{B} & \colhead{C} & \colhead{D} & \colhead{(dex)}\\
\colhead{(1)} & \colhead{(2)} & \colhead{(3)} & \colhead{(4)} & \colhead{(5)} & \colhead{(6)} & \colhead{(7)} & \colhead{(8)} & \colhead{(9)} & \colhead{(10)}}
\startdata
J0000+554 & 7 & 15.9 & -0.93 & 23.2 & \nodata & \nodata & \nodata & \nodata & \nodata \\ 
J0003-173 & 17 & 18.9 & -0.79 & 24.8 & 1.31 & -0.75 & \nodata & \nodata & 0.084 \\ 
J0004+131 & 5 & 16.6 & -0.88 & 15.8 & \nodata & \nodata & \nodata & \nodata & \nodata \\ 
J0006-000 & 29 & 27 & -0.76 & 36.6 & 1.43 & -0.7 & \nodata & \nodata & 0.142 \\ 
J0017+792 & 23 & 23.9 & -0.86 & 36.3 & 1.57 & -0.87 & \nodata & \nodata & 0.139 \\ 
J0019-124 & 5 & 18.1 & -0.89 & 30.1 & \nodata & \nodata & \nodata & \nodata & \nodata \\ 
J0020+154 & 28 & 37.4 & -1.13 & 58.5 & 1.61 & -1.04 & \nodata & \nodata & 0.08 \\ 
J0025-293 & 16 & 28 & -0.8 & 32.6 & 1.56 & -0.86 & -0.11 & -20.49 & 0.025 \\
\enddata
\tablecomments{Col. (1): VLSS source name.  Col. (2): Number of frequencies with measured flux densities.  Col. (3): VLSS 74 MHz flux density.  Col. (4): Slope of power-law fit to data with $\nu>300$ MHz.  Col. (5): The flux density at 74 MHz extrapolated using the power-law fit to data with $\nu>300$ MHz.  Col. (6-9): The parameters from equation (1) derived using a fit to the data for all sources with $N_{\nu}\geq 8$; for sources where a simple linear fit gives a quantitatively better fit (see text), only the A and B parameters are listed.  Col. 10: The rms difference between the observed and fitted values of $Y$ (equal to log $F_{\nu}$; see equation (1) and text).}
\label{fittab}
\end{deluxetable}

\clearpage

\end{document}